\documentclass[5p,times]{elsarticle}

\usepackage{verbatim}
\usepackage{hyperref}
\usepackage{multirow}
\usepackage{graphicx}
\graphicspath{ {D:/Latex} }

\begin{document}

\begin{frontmatter}

\title{Collatz-Weyl Generators: High Quality and High Throughput Parameterized Pseudorandom Number Generators}

\author[Independent researcher]{Tomasz R. Dziala}
\ead{CollatzGenerators@gmail.com}

\begin{abstract}
	We introduce the Collatz-Weyl Generators, a family of uniform pseudorandom number generators (PRNGs) which are based on generalized Collatz mappings, derived from the Collatz conjecture and Weyl sequences. The high-quality statistical properties of our generators is demonstrated by the fact that they pass stringent randomness tests used by the research and standardization community. The proposed Collatz-Weyl Generators have a number of important properties, including solid mathematical foundations, enablement of high throughput and low latency implementation, small code and/or ASIC size, enablement of producing multiple independent streams and potential of support of cryptographic applications.
\end{abstract}


\end{frontmatter}

\section{INTRODUCTION}
It is well known that no arithmetical method is available to automatically generate random numbers. The pseudorandom number generator (PRNG) is a deterministic algorithm that generates numbers that approximate to the properties of random numbers. Designing a good pseudorandom number generator is always a bit of an art, not strict mathematics. As von Neumann said, {\textquotedblleft}we are dealing with mere {\textquoteleft}cooking recipes{\textquoteright} for making digits; they can probably not be justified, but should merely be judged by their results{\textquotedblright} \cite{von1951various}.

Despite that fact, there are several mathematical features of a pseudorandom number generator that directly relate to the confidence we can have in it. The fundamental mathematical characteristics of a high-quality and reliable PRNG are a well defined period and the uniformity of the outputs generated by it. However, many applications require independent stream generation through the choice of a particular parameter or a jump ahead in the given sequence (cycle-splitting). In the era of microprocessor chips with multiple CPU cores that can simultaneously execute multiple threads, the scalability of the PRNG is an increasingly important feature as it renders it suitable for the generation of parallel pseudorandom numbers.

\bigskip

In this paper, the author proposes a scheme to create high-quality PRNGs. The proposed scheme is based on functions with strong mixing properties, called the generalized Collatz functions (GCFs). To eliminate the problem of short cycles we cmobine them with Weyl sequences. We focus on simplified GCFs to maximize efficiency in implementations of this scheme. However, we think that more complex GCFs are also suitable for the generation of pseudorandom numbers, especially in cryptography, owing to a wide parameter space and the non-invertibility\footnote{Note that when we use the term {\textquotedblleft}non-invertibility,{\textquotedblright} we refer only to graph structures of the state space of the generator. For instance, there may be several paths to reverse the state of such a generator. We do not refer to non-invertibility in the cryptographic sense at any point in this paper. However, we suggest that the non-cryptographic non-invertibility of some of the presented schemes may remain consistent with non-invertibility in the cryptographic sense. We leave this issue as a hypothesis, and do not analyze it in this paper.} of such mappings as this may result in a resistance to backtracking in specific schemes \cite{fischer2012backtracking}. Furthermore, past research has shown that such functions can be associated with factorization as well as the discrete logarithm problem, where this makes them suitable for cryptographic applications \cite{thakkar2016additive}. We leave this issue open for future research.

\section{COLLATZ-WEYL GENERATORS}

Collatz-Weyl Generators (CWGs) are chaotic generators that are combined with Weyl sequences to eliminate the risk of short cycles. They have a large period, a uniform distribution, and the ability to generate multiple independent streams by changing their internal parameters (Weyl increment). CWGs owe their exceptional quality to the arithmetical dynamics of non-invertible, generalized, Collatz mappings based on the well-known Collatz conjecture. There is no jump function, but each odd number of the Weyl increment initiates a new unique period, which enables quick initialization of independent streams.

\subsection{Generalized Collatz Sequences}

Generalized Collatz sequences have been considered by several researchers. Conway \cite{conway1972unpredictable} and, subsequently, Kurtz and Simon \cite{kurtz2007undecidability} considered the following generalizations (generalized Collatz functions):

\newtheorem{definition}{Definition}
\begin{definition}\label{defn:1}
A function g is called a Collatz function if there is an integer m together with non-negative rational numbers $\{a_{i}, b_{i}:i<m\}$ such that whenever $x \equiv i \pmod m$, then $g(x) = a_{i} \cdot x+b_{i}$ is an integral.
\end{definition}

This is a natural generalization: The function of the Collatz problem assumes this form for $m=2$, $a_{0}=\frac{1}{2}$, $b_{0}=0$, $a_{1}=3$, and $b_{1}=1$. We consider similar but even more general functions. The main modification in the functions that we use is the introduction of the modulo operation. That is, instead of

\begin{center}
  $g(x) = a_{i} \cdot x + b_{i}$,
\end{center}

we compute:

\begin{center}
  $g(x) = a_{i} \cdot x + b_{i} \pmod {p}$
\end{center}

where $p$ is a natural number. However, it is a power of 2 in most of the cases considered.

\bigskip

Let $j$ be any natural number and $x$ be an integer. We then consider functions of the form:
\begin{equation}
f(x)=a_{i_{j}} \cdot x^{j} + a_{i_{j-1}} x^{j-1} + \ldots + a_{i_{2}} \cdot x^{2} + a_{i_{1}} \cdot x + a_{i_{0}} \pmod{p_{i}}
\end{equation}

with rational coefficients $a_{i_{0}},\ldots,a_{i_{j}}$, where $p_{i}$ may be any natural number.

Let $h(x)$ be a polynomial of the form:
\begin{equation}
h(x)=b_{i_{j}} \cdot x^{j} + b_{i_{j-1}} x^{j-1} + \ldots + b_{i_{2}} \cdot x^{2} + b_{i_{1}} \cdot x + b_{i_{0}} \pmod {q_{i}}
\end{equation}

with rational coefficients $b_{i_{0}},\ldots,b_{i_{j}}$, where $q_{i}$ may be any natural number.

\begin{definition}\label{defn:2}
A function $f$ is called a generalized Collatz function if there is an integer m, along with rational numbers \newline
$\{a_{i_{0}},\ldots,a_{i_{j}},b_{i_{0}},\ldots,b_{i_{j}} \colon i<m\}$ and natural numbers $\{p_{i},q_{i} \colon i<m\}$, such that whenever $h(x) \equiv i \pmod m$, then $f(x)$ is an integral.
\end{definition}

All of the arithmetic operations of $f$ may be performed in Galois fields, or as carry-less multiplication and addition.

In some representations, the coefficients  $a_{i_{0}},\ldots,a_{i_{j}}$, $b_{i_{0}},\ldots,b_{i_{j}}$ must be odd integers. In these cases, we can apply the following operation on even coefficients: 
operation - $a_{i_{0}} \mathbin{|} 1,...,a_{i_{j}} \mathbin{|} 1,b_{i_{0}} \mathbin{|} 1,...,b_{i_{j}} \mathbin{|} 1$, where $\mathbin{|}$ is the bitwise OR. This operation converts even coefficients into odd coefficients; when it is applied to the latter, they remain odd.

In some representations, the variables and coefficients of $f(x)$ and $h(x)$, as well as all the polynomials, may be shifted by a certain number of bits by using logical or circular bit shifts, and the addition operation may be replaced by the bitwise XOR operation.

\subsection{General Scheme}

Definition~\ref{defn:2} allows for the creation of a wide class of State Functions that can be implemented in PRNGs. However, because such functions usually define non-invertible pseudorandom mappings, we mix their states with Weyl sequences--counters modulo $2^{w}$, where each increment is an odd number \cite{marsaglia2003xorshift,widynski2017middle}--to provide the generator with a guaranteed minimum period. An additional step involving mixing coefficients and the outputs is also usually required to yield results of better quality.

Therefore, we define the general algorithm of Collatz-Weyl Generators as follows:

\begin{enumerate}
\item Mix \texttt{coefficients}.
\item Compute \texttt{$f$}.
\item Combine the state of the generator with the \texttt{Weyl sequence}.
\item Return scrambled \texttt{state}.
\end{enumerate}

\subsection{Algorithms}

We now define some of the most efficient and the simplest implementations of Collatz-Weyl Generators: the CWG128, CWG64, and CWG128-64 generators. To render Collatz-Weyl Generators more efficient, we consider the simplest GCFs with $m=1$, where this helps avoid conditional branching in the code. Because $h(x) = x$ and $m=1$, the algorithm always executes the same instruction as $x \equiv 0 \pmod 1$. Therefore, there is only one coefficient $a_{0_{0}}$, which we consider to be a $a$ variable in our code.

\begin{figure*}
\center
\includegraphics[scale=0.8]{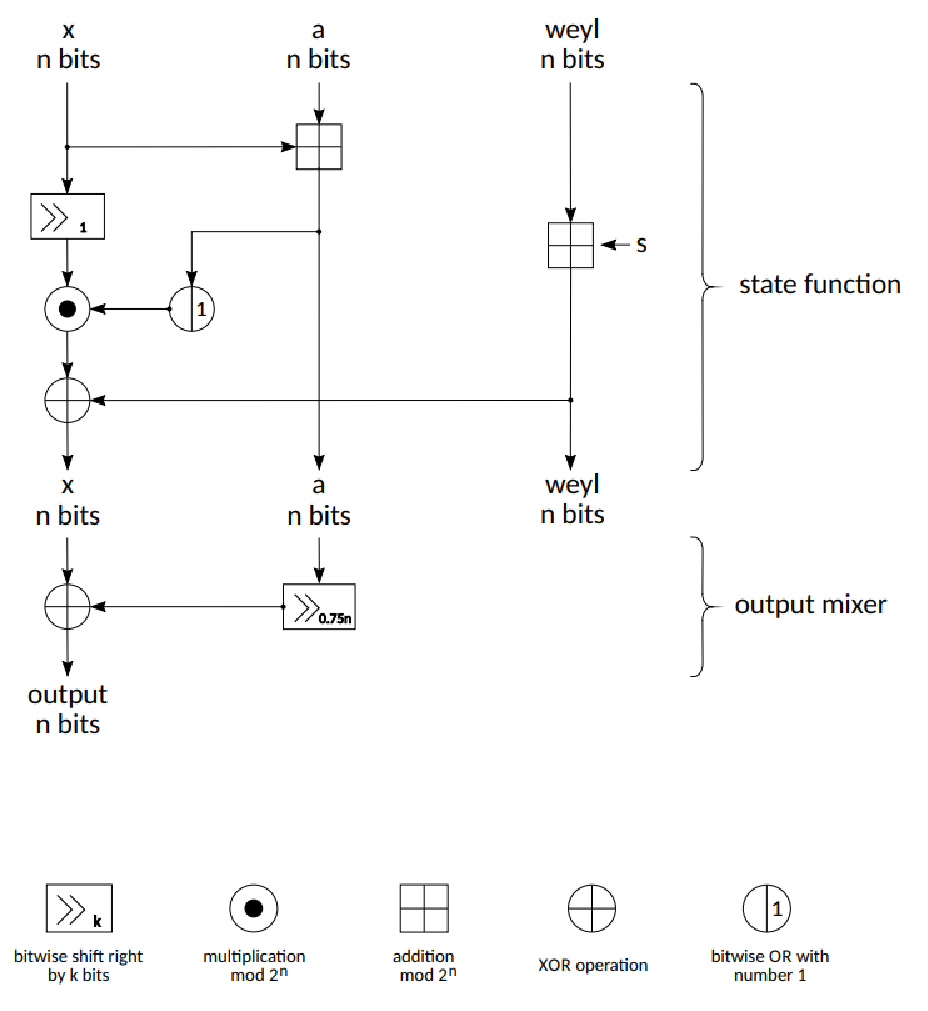}
\caption{Block diagram of n-bit Collatz-Weyl Generator.\label{fig:1}}
\end{figure*}

Figure \ref{fig:1} shows the general scheme of the n-bit Collatz-Weyl Generator. Figures \ref{fig:2}, \ref{fig:3}, and \ref{fig:4} show the specific implementations of the generation operations of the CWG64, CWG128-64, and CWG128 generators in C++. At first glance, these generators do not resemble the original Collatz function:

\bigskip

\begin{center}
$
C(x) = 
\left\{
\begin{array}{cl}
\frac{3x+1}{2} & \mbox{if}~ x \equiv 1~~ (\bmod ~2 ) , \\
~~~ \\
\frac{x}{2} & \mbox{if} ~~x \equiv 0~~ (\bmod~2) .
\end{array}
\right.
$
\end{center}

\bigskip

However, we can write Collatz function as follows by using bitwise operations:

\bigskip

\begin{center}
  $
C(x) = 
\left\{
\begin{array}{cl}
x = x + ((x + 1) \gg 1) & \mbox{if}~ x \equiv 1~~ (\bmod ~2 ) , \\
~~~ \\
x = x \gg 1 & \mbox{if} ~~x \equiv 0~~ (\bmod~2) .
\end{array}
\right.
$
\end{center}

\bigskip

The main types of transformations performed by the original Collatz function are bitwise shifts of the state by $1$ bit or the additions of a number to it followed by a bitwise shift. All three proposed Collatz-Weyl Generators above also contain this transformation, which induces arithmetical chaos when combined with multiplication. Using bitwise OR with the number 1 on one of the factors gives slightly better statistical results, according to our experiments. In each generator, the lower bits of $x$ are XORed with the upper bits of $a$ in the output scrambler, due to the fact that the lower bits of $x$ do not have sufficient statistical quality.

\begin{figure*}
\center
\begin{verbatim}
static uint64_t x, a, weyl, s; // s must be odd

uint64_t CWG64(void){
    x = (x >> 1) * ((a += x) | 1) ^ (weyl += s);
    return a >> 48 ^ x;
}
\end{verbatim}
\caption{Implementation of the "generate{\textquotedblright} operation of the CWG64 generator in C++.\label{fig:2}}
\end{figure*}

\begin{figure*}
\center
\begin{verbatim}
static __uint128_t x;
static uint64_t a, weyl, s; // s must be odd

__uint128_t CWG128_64(void){
    x = (x | 1) * ((a += x) >> 1) ^ (weyl += s);
    return a >> 48 ^ x;
}
\end{verbatim}
\caption{Implementation of the {\textquotedblleft}generate{\textquotedblright} operation of the CWG128-64 generator in C++.\label{fig:3}}
\end{figure*}

\begin{figure*}
\center
\begin{verbatim}
static __uint128_t c[4]; // c[0] must be odd

__uint128_t CWG128(void){
    c[1] = (c[1] >> 1) * ((c[2] += c[1]) | 1) ^ (c[3] += c[0]);
    return c[2] >> 96 ^ c[1];
}
\end{verbatim}
\caption{Implementation of the {\textquotedblleft}generate{\textquotedblright} operation of the CWG128 generator in C++.\label{fig:4}}
\end{figure*}

\newpage

\section{PERIOD}
Collatz-Weyl Generators can be considered to be a combination of two sequences: one defined by a non-invertible pseudorandom mapping, and the other being a Weyl sequence of an irreducible, minimum-period length. The cycle length of a typical combined generator is the least common multiple of the cycle lengths of the individual generators \cite{fog2015pseudo}. This technique for extending the period of the generator has been known since Marsaglia \cite{marsaglia2003xorshift}, and was used by Widynski \cite{widynski2017middle}. However, the Marsaglia design involves two separate generators that are combined in the output. In our design here, the state of the generator, and not merely the output, is combined with the Weyl sequence in a similar manner to that in Widynski's middle-square Weyl generator. Therefore, there is no least common multiple of two separate generators because we are dealing with only one generator with a dual structure. However, the period of the Weyl sequence is well known, and we may estimate period of the non-invertible pseudorandom mapping, to determine the overall period of the generator. Because the period of an n-bit Weyl sequence is irreducible, the minimum period of every CWG generator is guaranteed to be $2^{n}$. The period of the non-invertible pseudorandom sequence may be approximated from the theory of random mappings \cite{flajolet1989random}. The expected path to reach the cycle and the expected cycle length are given by the same formula $$\sqrt{\frac{\pi}{8} \cdot 2^{n}}$$ for an n-bit state generator. On the whole, we start to see repeated patterns after $$2 \cdot \sqrt{\frac{\pi}{8} \cdot 2^{n}}$$ iterations. If we had a pseudorandom number generator composed of two 64-bit substates based on a non-invertible pseudorandom sequence, then we would calculate the number of iterations after which we expect repetitions by substituting $n=128$. But here we must also consider the irreducible substate $weyl$. In CWG64 generator, only the non-invertible part of the state $x$ is xored with 64-bit $weyl$ substate, and $a$ is not. Hence, only 64 bits of the generator state will preserve the non-invertible pseudorandom mapping properties, while the other 64 bits would follow the period of the Weyl sequence. Therefore, the generator period will be a product of the Weyl sequence period and the expected value of path and cycle length of the 64-bit non-invertible random mapping: $$\sqrt{\frac{\pi}{8} \cdot 2^{64}} \cdot 2^{64}$$. Hovewer, this does not imply that $x$ will have exactly the $weyl$ period, nor that $a$ will follow the period predicted by pseudorandom mapping statistics, as they are both mixed together. Their periods will be given by the same formula and synchronized, especially because substate $a$ is simply the modulo sum of $x$. Consequently it will have a path to the cycle of the same length as $x$ and a cycle perfectly synchronized with $x$. Thus, in overall we expect to see repeats in CWG64 generator after $$2 \cdot \sqrt{\frac{\pi}{8} \cdot 2^{64}} \cdot 2^{64} \sim 2^{96.33}$$ numbers. By analogy, we start to see repeats in the CWG128-64 generator and the CWG128 generator after $$2 \cdot \sqrt{\frac{\pi}{8} \cdot 2^{64}} \cdot 2^{64} \sim 2^{96.33}$$ and $$2 \cdot \sqrt{\frac{\pi}{8} \cdot 2^{128}} \cdot 2^{128} \sim 2^{192.33}$$ generated numbers, respectively. Note that while CWG128-64 has a larger state than CWG64, the high 64-bits of $x$ have no effect on other bits of the state of the generator, and therefore have no
effect on the period. The period is thus no different from that of the CWG64 generator.

Because the $2^{n}$ cycle length of the n-bit Weyl sequence is guaranteed, a cycle shorter than the length of the Weyl sequence cannot exist in our generator. Every cycle will be a multiple of the Weyl sequence's cycle length. However, arbitrarily short paths to reach the cycles may occur. In particular, we can imagine an initialization that starts the generator in a certain cycle, in which case the length of the path to reach the cycle would be zero. A generator initialized with the same $s$ or $c[0]$ can have several disjoint cycles. The expected number of components (in our case, this corresponds to the number of cycles, since each component contains one unique cycle) based on the theory of non-invertible random mappings \cite{flajolet1989random} is given by: $$\frac{\ln(2^{n})}{2}$$, where $n$ is equal to $64$ for CWG64 and CWG128-64, and $128$ for CWG128.

\bigskip

In summary, the minimum guaranteed period of CWG64 under every uniquely initialized Weyl sequence (there are $2^{63}$ such initializations) is $2^{64}$, and we expect to see repeats after $\sim 2^{96.33}$ generated numbers. That of CWG128-64 under every uniquely initialized Weyl sequence (there are $2^{63}$ such initializations) is $2^{64}$, and we expect to see repeats after $\sim 2^{96.33}$ generated numbers. The minimum guaranteed period of CWG128 under every uniquely initialized Weyl sequence (there are $2^{127}$ such initializations) is $2^{128}$, and we expect to see repeats after $\sim 2^{192.33}$ generated numbers.

\bigskip

Because it is possible for the generator to fall into a much shorter path to reach the cycle or/and into a much shorter cycle than expected, we do not recommend using it beyond the minimum guaranteed period. Every generator also has an internal bias that may lead to a difference between the actual statistics of the pseudorandom random mappings and those related to true random mappings.

\section{UNIFORMITY}

Uniformity is a property that implies that a PRNG returns each value in the range of possible results with equal probability. The Weyl sequence provides not only a minimum guaranteed period, but also uniformity. Widynski gave simple proof of this \cite{widynski2017middle}.

\newtheorem{theorem}{Theorem}
\begin{theorem}
  Let $x$ be a random but not necessarily uniform bit stream, and let $weyl$ be a uniform but not necessarily random bit stream. Then, $x \oplus weyl$ is a uniform bit stream.
  \end{theorem}


Let us consider n-bit quantities. We can demonstrate uniformity by showing that the probability of any given output is $\frac {1}{2^{n}}$. 


\newproof{proof}{Proof}
\begin{proof}
Given an arbitrary element $y$ in $[0, 2^{n}-1]$, we compute the probability of $y$, where $y = x \oplus weyl$.
$$P(y) = \sum_{x=0}^{2^{n}-1} (P(x) \cdot P(weyl))$$ \newline
Because $weyl$ is uniform, $P(weyl)$ is $\frac {1}{2^{n}}$.  Thus, 
$$P(y) = \sum_{x=0}^{2^{n}-1} (P(x) \cdot \frac {1}{2^{n}})$$ \newline,
which is equivalent to
$$P(y) = \frac {1}{2^{n}} \sum_{x=0}^{2^{n}-1} (P(x))$$ \newline
As the sum of the probabilities of all possible outcomes is one, 

$$P(y) = \frac {1}{2^{n}}$$.

\end{proof} 

There are $2^{n}$ possible outputs in the n-state Collatz-Weyl Generator, each of which has a probability of occurrence of $\frac {1}{2^{n}}$. Such a generator produces uniformly distributed values of $y = x \oplus weyl$, and $a$ is thus also uniform as it is simply the modulo sum of uniform numbers. The output of Collatz-Weyl Generators is therefore a bitwise shift and an XOR operation on two uniform variables, and thus is also uniform. Note that this proof refers only to the occurrence of each value in the range of possible results with an equal probability. This does not mean that each value in the range of $2^{n}$ outputs occur exactly once. Furthermore, even if our generator were based on a perfectly random bit stream, we expect a particular number of repetitions in the values of the generated numbers due to the birthday problem \cite{wiki:birthday}.

\section{BIRTHDAY PROBLEM}


As we show below, Collatz-Weyl Generators pass statistical tests above the guaranteed minimum period. However, because they are based on pseudorandom mappings, there is no risk of flaws due to the birthday problem such that they can be used in the range of the full period, rather than its square root as in the case of typical PRNGs \cite{caflisch1998monte, larcher1998random, gentle2003random, o2015pcg}. The uniformity of the generators proven in the previous section does not guarantee that all outputs will have occurred exactly the same number of times once a generator has completed a full period, but implies only that a generator returns each value in the range of possible results with an equal probability. However, some numbers may appear twice or more by chance, as in case of the generation of true random numbers. Even if we initialized a generator so that it did not go through a path to reach the cycle but started at a particular cycle (in the worst case, such a cycle may have only the length of the underlying Weyl sequence), numbers in this cycle usually do not form a one-cycle permutation as in the LCG generator, for example, in which each number occurs exactly once in a period (which leads to a deviation from the true randomness detectable by the birthday test). We expect cycles composed of random numbers in CWGs that are chosen uniformly from the range of possible outputs, and therefore some numbers may repeat. Of course, we must remember that each pseudorandom number generator has an internal bias, and the proposed generators simulate only randomness. However, PRNGs based on random mappings seem to be the closest we can approach to emulating true randomness. This holds as long as they do not contain systematic bias, and provide guarantees on the fundamental characteristics of the generator related to the confidence that we can have in it, such as its period and uniformity.

\section{MULTIPLE INDEPENDENT STREAMS}

As Collatz-Weyl Generators are chaotic generators, every unique initialized Weyl sequence gives a separate and distinctive stream. There is thus no risk of overlap. This holds even if we consider two states of two Collatz-Weyl Generators that would be equal at some point:

\bigskip

\begin{center} $x_{1} = x_{2}$
  \end{center}
  
\bigskip

\begin{center}
  $a_{1} = a_{2}$
  \end{center}

\bigskip

\begin{center}
  $weyl_{1} = weyl_{2}$
  \end{center}
  
\bigskip

We define two streams with two kinds of increments in the value of $s$ such that $s_{1} \neq s_{2}$ needs to diverge. Therefore the next states of the Weyl sequences,

\bigskip

\begin{center} $weyl_{1} + s_{1} \neq weyl_{2} + s_{2}$,
  \end{center}
  
\bigskip

differ, as do all the future states of the generator.

\subsection{Initialization}

The increment $s$ must be odd to ensure the maximum period of the Weyl sequence. All the other variables can be initialized by using any number. We suggest seeding $s$ by an odd number and setting all other elements to zero for the sake of simplicity, in single-stream applications.

In multi-stream applications, to ensure the complete independence of the $2^{64}$ streams, the first 48 and 96 states of each stream should be skipped in the CWG64 and CWG128-64, and the CWG128 generators, respectively. Otherwise, minor correlations may occur in the first outputs of the successively initialized streams, especially if they are initialized by using the numbers $1, 3, 5, ...$ or another counter. This is because CWG generators initialized with similar but not the same states need a certain number of iterations to decorelate. For this purpose, the user may call the initializer (Figures~\ref{fig:5}, \ref{fig:6}, and \ref{fig:7}). Every stream can be initialized with any 64-bit odd number by using the initializer().

\begin{figure*}
\center
\begin{verbatim}
void CWG64_initializer(void){
	for (int i = 0; i < 48; i++){
	x = (x >> 1) * ((a += x) | 1) ^ (weyl += s);
    }
}
\end{verbatim}
\caption{Implementation of the CWG64 initializer in C++.\label{fig:5}}
\end{figure*}

\begin{figure*}
\center
\begin{verbatim}
void CWG128_64_initializer(void){
	for (int i = 0; i < 48; i++){
	x = (x | 1) * ((a += x) >> 1) ^ (weyl += s);
    }
}
\end{verbatim}
\caption{Implementation of the CWG128-64 initializer in C++.\label{fig:6}}
\end{figure*}

\begin{figure*}
\center
\begin{verbatim}
void CWG128_initializer(void){
	for (int i = 0; i < 96; i++){
	c[1] = (c[1] >> 1) * ((c[2] += c[1]) | 1) ^ (c[3] += c[0]);
	}
}
\end{verbatim}
\caption{Implementation of the CWG128 initializer in C++.\label{fig:7}}
\end{figure*}

For faster initialization, the user may apply the Splitmix64 \cite{steele2014fast} and Splitmix63 generators 
(a modified Splitmix64 generator), and use their outputs to serially fill the values of $x$ and $s$. The implementations of the Splitmix64 and Splitmix63 generators in C++ are shown in Figures~\ref{fig:8} and \ref{fig:9}. We recommend using the method shown in Figure \ref{fig:10} to serially seed the state of CWG64, this makes it possible to initialize $2^{63}$ streams,  the method presented in Figure \ref{fig:11} to serially seed the state of CWG128-64, this makes it possible to initialize $2^{63}$ streams, and to serially seed the state of CWG128 using the method shown in Figure \ref{fig:12}, this makes it possible to initialize $2^{63}$ streams.

\begin{figure*}
\center
\begin{verbatim}
static uint64_t y; /* The state can be seeded with any value. */

uint64_t splitmix64() {
	uint64_t z = (y += 0x9e3779b97f4a7c15);
	z = (z ^ (z >> 30)) * 0xbf58476d1ce4e5b9;
	z = (z ^ (z >> 27)) * 0x94d049bb133111eb;
	return z ^ (z >> 31);}
\end{verbatim}
\caption{Implementation of the generate operation of the Splitmix64 generator in C++.\label{fig:8}}
\end{figure*}

\begin{figure*}
\center
\begin{verbatim}
static uint64_t y; /* The state can be seeded with any value. */

uint64_t splitmix63() {
	uint64_t z = (y += 0x9e3779b97f4a7c15) & 0x7fffffffffffffff;
	z = ((z ^ (z >> 30)) * 0xbf58476d1ce4e5b9) & 0x7fffffffffffffff;
	z = ((z ^ (z >> 27)) * 0x94d049bb133111eb) & 0x7fffffffffffffff;
	return z ^ (z >> 31);}
\end{verbatim}
\caption{Implementation of the generate operation of the Splitmix63 generator in C++.\label{fig:9}}
\end{figure*}

\begin{figure*}
\center
\begin{verbatim}
a = 0, weyl = 0;
x = splitmix64();
s = (splitmix63() << 1) | 1;
\end{verbatim}
\caption{Implementation of the initialization of the CWG64 generator by the Splitmix63 generator in C++.\label{fig:10}}
\end{figure*}

\begin{figure*}
\center
\begin{verbatim}
a = 0, weyl = 0;
x = ((__uint128_t)splitmix64() << 64) | splitmix64();
s = (splitmix63() << 1) | 1;
\end{verbatim}
\caption{Implementation of the initialization of the CWG128-64 generator by the Splitmix64 and Splitmix63 generators in C++.\label{fig:11}}
\end{figure*}

\begin{figure*}
\center
\begin{verbatim}
c[1] = splitmix64();
c[0] = ((__uint128_t)splitmix64() << 64) | ((splitmix63() << 1) | 1);
\end{verbatim}
\caption{Implementation of the initialization of the CWG128 generator by the Splitmix64 and
Splitmix63 generators in C++.\label{fig:12}}
\end{figure*}

The Splitmix63 generator is required to obtain a unique and non-repetitive sequence of 63-bit numbers that are needed to initialize odd Weyl increments. This would not be possible with the Splitmix64 generator\footnote{There is a low probability of obtaining identical numbers, i.e., a collision, whereby 64-bit numbers are converted into 63-bit numbers.}. Research has shown \cite{matsumoto2007common} that initialization performed by using a generator that is radically different from the one initialized enables the avoidance of a correlation on similar seeds. A similar method has been used in the Xoroshiro family of PRNGs \cite{blackman2021scrambled}. We think that this will also work for our generators.

As Collatz-Weyl Generators can generate multiple independent streams, they can be easily parallelized.

\section{STATISTICAL TESTS}

All the proposed generators passed the NIST test suite, Dieharder, TestU01 (Big Crush), and PractRand up to 8 TB \cite{rukhin2001statistical,brown2018dieharder,l2007testu01,doty10practically}. Since standard versions of CWG generators may have to big states to fail and so applied tests would not be able to identify their limitations \cite{too-big-to-fail,pcg-passes-practrand,o2015pcg}, we examined scaled down versions of CWG64 (CWG32 and CWG16) and CWG128-64 (CWG64-32 and CWG32-16). All scaled-down versions passed PractRand above the limit of their minimum guaranteed periods, which confirms the quality of their full-scale versions.

\subsection{Methodology}

We executed 200 complete runs of the three proposed generators to verify the dependence of random
numbers from the same sequence (intra-stream correlation), and 400 complete runs to verify their inter-stream correlations \cite{salmon2011parallel,tan2021thundering}.

We used two methods to verify the dependence between the separate sequences. The first was adapted from work by Li et~al. \cite{li2013software}. We first interleaved 4, 8, 128, and 1024 sequences into a single sequence and then evaluated the resulting sequence. 100 complete runs were performed for initialization by the initalizator() and 100 complete runs for initialization by the Splimixes. In the second method we tested consecutive firsts, seconds, fourths, and 10ths of the outputs (so we created one stream composed of the first outputs of serially initialized generators, then tested a stream composed of the second outputs of serially initialized generators, and so on). 100 complete runs for initialization by the initializer() and 100 complete runs for initialization by the Splimixes were performed.

\subsection{Results}

Tables \ref{tab:I} and \ref{tab:II} show the results of tests of the CWG64, CWG128-64, CWG128 and scaled-down versions of CWG generators on the NIST test suite, Dieharder, TestU01, and PractRand.

\bigskip 

\begin{table*}[ht]
\centering
\caption{Statistical tests on the CWG generators for intra-stream correlations.\label{tab:I}}
\begin{tabular}{c c c c c c}
\hline
\multirow{5}{*}{Algorithm} & \multirow{5}{*}{Min. period (bits)} \\
\multicolumn{5}{l}{$\quad \ \ \ \ \ \qquad \qquad \qquad \qquad \qquad \qquad \qquad \qquad \qquad \qquad$ Intra-stream correlation} \\[6pt]
\cline{3-6}
 &   & \multirow{2}{*}{NIST} & \multirow{2}{*}{Dieharder} & \multirow{2}{*}{TestU01} & \multirow{2}{*}{PractRand (bits)} \\[11pt]
\hline \\
\textbf{CWG128}  & $2^{135}$  & $>$ min. pass rate    &passed  & passed    & $>2^{46}$ \\[7pt]
\textbf{CWG128-64} & $2^{71}$ & $>$ min. pass rate    &passed   & passed    & $>2^{46}$ \\[7pt]
\textbf{CWG64}  & $2^{70}$   & $>$ min. pass rate    &passed   & passed    & $>2^{46}$ \\[7pt]
CWG32    & $2^{37}$ & $>$ min. pass rate    &not applied  & not applied    & $< 2^{44}$ \\[7pt]
CWG64-32   & $2^{38}$ & $>$ min. pass rate    &not applied  & not applied    & $< 2^{45}$ \\[7pt]
CWG16    & $2^{20}$  & not applied    &not applied  & not applied    & $< 2^{29}$   \\[7pt]
CWG32-16  & $2^{21}$ & not applied    &not applied  & not applied  & $< 2^{25}$   \\[7pt]
\hline
\end{tabular}
\end{table*}

\begin{table*}[ht]
\centering
\caption{Statistical tests of CWG generators for inter-stream correlations.\label{tab:II}}
\begin{tabular}{c c c c c c}
\hline
\multirow{5}{*}{Algorithm} & \multirow{5}{*}{No. of streams} \\
\multicolumn{5}{l}{$\quad \ \qquad \qquad \qquad \qquad \qquad \qquad \qquad \qquad \qquad \qquad$ Inter-stream correlation} \\[6pt]
\cline{3-6}
 &   & \multirow{2}{*}{NIST} & \multirow{2}{*}{Dieharder} & \multirow{2}{*}{TestU01} & \multirow{2}{*}{PractRand (bytes)} \\[11pt]
\hline \\
\textbf{CWG128}  & $2^{127}$   & $>$ min. pass rate    &passed  & passed    & $>2^{43}$, $>2^{43}$  \\[7pt]
\textbf{CWG128-64} & $2^{63}$  & $>$ min. pass rate    &passed   & passed    & $>2^{43}$, $>2^{43}$  \\[7pt]
\textbf{CWG64}  & $2^{63}$ & $>$ min. pass rate    &passed   & passed    & $>2^{43}$, $>2^{43}$ \\[7pt]
CWG32    & $2^{31}$  & not applied    &not applied  & not applied    & $> 2^{43}$, $< 2^{34}$ \\[7pt]
CWG64-32   & $2^{31}$  & not applied    &not applied  & not applied    & $> 2^{43}$, $< 2^{35}$  \\[7pt]
CWG16    & $2^{15}$ & not applied    &not applied  & not applied    &$< 2^{24}$, $< 2^{17}$  \\[7pt]
CWG32-16  & $2^{15}$  & not applied    &not applied  & not applied  & $< 2^{20}$, $< 2^{18}$ \\[7pt]
\hline
\end{tabular}
\end{table*}%
\raggedbottom

Table \ref{tab:I} shows that the Collatz-Weyl Generators passed all the single-stream tests, and this confirms their excellent statistical quality.
Because the minimum guaranteed period of the CWG32 and CWG64-32 generators was $2^{37}$ and $2^{38}$ bits while they passed the tests up to $2^{43}$ and $2^{44}$ bits, they can generate numbers up to a threshold that is 64 times higher than is required by the minimum guaranteed period without incurring failure. The period of the CWG16 and CWG32-16 generators was $2^{20}$ and $2^{21}$ bits while they passed the tests up to $2^{28}$ and $2^{24}$ bits (which is consistent with the expected path to the cycle and the cycle length). Therefore, they can generate numbers up to a threshold 8-256 times higher than is required by the minimum guaranteed period without incurring failure. Hence, we expect that larger-scale versions of these generators would also pass the tests above the threshold of the minimum guaranteed period. NIST, Dieharder, and TestU01 were not applied to the 16-bit versions because they generated too small a number of bits to pass these tests.

\bigskip

Table \ref{tab:II} shows that the Collatz-Weyl Generators passed all tests of inter-stream correlations, and this confirms that they can be used to generate multiple independent sequences. The average results of the streams initiated by the Splitmixes (Figures~\ref{fig:5}, \ref{fig:6}, and \ref{fig:7}) or by the random numbers in the case of scaled down CWGs are on the left side of the PractRand column, and the results of the streams initiated by the counter and initializer() (Figures~\ref{fig:10}, \ref{fig:11}, and \ref{fig:12}) are on the right side. In the reduced versions, the results are given per stream, for example, if 128 streams were interleaved, the result is divided by 128.

NIST, Dieharder, and TestU01 were not applied to the scaled-down versions because initialization by using the counter limited the outputs of the CWG64-32 and CWG32 generators to only $2^{31}$ numbers, and the CWG32-16 and CWG16 generators to $2^{15}$ numbers. The size of these data was too small for the generators to have passed these tests. However, we applied PractRand to the CWG64-32 and CWG32 generators for up to $2^{31}$ streams, and to the CWG32-16 and CWG16 generators for up to $2^{15}$ streams. They passed the test until the counter had looped.

\bigskip

The results of the above tests suggest that the proposed generators can be used in the full range of the guaranteed period and parallelized on a large scale without incurring the risk of failure. $2^{64}$ generated numbers can be used in each of the $2^{63}$ streams of the CWG64 and CWG128-64 generators, and $2^{128}$ generated numbers can be used in each of the $2^{127}$ streams of the CWG128 generator.


\subsection{Birthday Test}

Additionally, we performed 1000 runs of the birthday test on a scaled-down version of the Collatz-Weyl Generator CWG32 and 32768 birthday tests on version CWG16 \cite{birthdaytest}. First we calculated how many outputs to look at and how many repeats we expect to see, according to an estimate of the expected number of repeats based on a binomial distribution: $$r \approx n - d \left( 1 - \left( 1 - \frac{1}{d} \right)^{n} \right)$$, where d is the number of possible different outputs that the generator can produce, and n is the number of samples; then we collected outputs, and counted how many repeats we saw. Finally we calculated p-values for the obtained results. A birthday test can cause some popular PRNGs to fail quickly because non-zero repetitions are expected in a true random number generator over its period of operation. On the contrary, the numbers generated by most PRNGs cannot be repeated in the range of their period. This is not the case with Collatz-Weyl Generators. The average p-value calculated based on observed repetitions in 1000 runs of the CWG32 version, initalized with random seed values, for the birthday test was $0.507$. In the case of CWG16 we tested all possible $2^{15}$ seeds $s$ (every odd $s$, with other variables set to $0$) in the full range of minimum guaranteed period ($2^{16}$ outputs). In an ideal random number generator, the average expected number of duplicates in the range of $2^{16}$ 16-bit numbers is 24109.2, while the average number of duplicates in all streams of the CWG16 generator was 24108.4, and the average p-value was $0.500$. These results are therefore indistinguishable from true random number generation under the birthday test and confirm that the Collatz-Weyl Generators can be used in the full range of their minimum guaranteed period without incurring the risk of failing that test.

\section{COMPARATIVE TESTS OF THROUGHPUT}

We measured the time taken by the generators to generate $10^9$ random numbers by using an AMD Ryzen 5 4600H, 3 GHz processor with an Ubuntu Windows Subsystem for Linux (WSL) with gcc
version 9.4.0. We also assessed the performance of several popular PRNGs for the sake of comparison. Ranlux++ and MixMax17 were run on a MacBook Pro 2.6 GHz, Intel Core i7 CPU in the ROOT library developed by CERN \cite{ROOT}. Table \ref{tab:III} lists the results.

\bigskip

\begin{table*}[ht]
\centering
\caption{Assessing the throughput of CWGs and several popular PRNGs.\label{tab:III}}
\begin{tabular}{c | c | c | c | c}
\hline
Algorithm & Footprint (bits) & Period & ns/64 bits & cycles per byte \\[7pt]
\hline
\textbf{CWG128-64}   & 256   & $>2^{64}$   & 0.59    & 0.22  \\

\textbf{CWG128}  & 384   & $>2^{128}$   & 0.77    & 0.29  \\

\textbf{CWG64}   & 192    & $>2^{64}$  & 1.12    & 0.42  \\

xoroshiro128++   & 128    & $2^{128}$  & 0.88    & 0.33  \\

xoshiro256++   & 256    & $2^{256}$  & 0.96    & 0.36  \\

xoshiro512+   & 512    & $2^{512}$  & 1.01    & 0.38  \\

LCG64   & 128    & $2^{128}$  & 0.93    & 0.35  \\

PCG-DXSM   & 128    & $2^{128}$  & 1.09    & 0.41  \\

MWC256   & 256    & $\sim 2^{255}$  & 0.83    & 0.31  \\

GMWC256   & 256    & $\sim 2^{255}$  & 1.39    & 0.52  \\

SFC64   & 256    & $2^{64}$  & 0.93    & 0.35  \\

SplitMix64   & 64    & $2^{64}$  & 0.72    & 0.27  \\

Ranlux++   &  1152   & $\sim 2^{576}$  & 12.12    & 4.55  \\  

MixMax17   & 1088    & $\sim 2^{977}$  & 5.20    & 1.95  \\

MT19937-32   & 20032    & $2^{19937}-1$  & 1.76   & 0.66 \\

Philox   & 384    & $2^{256}-1$  & 3.28    & 1.23  \\
\hline
\end{tabular}
\end{table*}%

Table \ref{tab:III} shows that the CWGs outperformed the other designs. However some of the generators considered here were not comparable to Collatz-Weyl Generators because they encountered statistical issues.

In addition to their simple and lightweight code, an important advantage of variants of Collatz-Weyl Generators that use 128-bit integers is that they output 128 bits per call. This makes them efficient in terms of the number of bytes produced per unit time.\vspace*{7pt}

\section{CONCLUSION AND FUTURE RESEARCH}

Collatz-Weyl Generators provide a bridge between an unsolved class of problems based on the generalized Collatz functions and the generation of pseudorandom numbers. Our key contribution to practical applications of generalized Collatz functions is the elimination of the problem of divergence to infinity of sequences based on these functions, through the use of the modulo operator, and the problem of short cycles that often characterize such sequences, by combining them with Weyl sequences.

We have not presented and examined all Collatz-Weyl Generators that can be derived from the scheme proposed in this study, but have investigated only those that we found to be the simplest and most efficient. Perhaps some Collatz-Weyl Generators still await discovery. An interesting issue to explore in this context is the cryptographic potential of Collatz-Weyl Generators.

\bigskip

\begin{figure*}
\center
\begin{verbatim}
static __uint128_t x, weyl, s; // s must be odd 

bool Collatz_Weyl(void){
    x = (-(x & 1) & (x + ((x + 1) >> 1)) | ~-(x & 1) & (x >> 1)) ^ (weyl += s);
    return x & 1;
}
\end{verbatim}
\caption{Constant-time implementation of the Collatz-Weyl Generator based on the original "3x+1" sequences in C++ by using bitwise and logical operations.\label{fig:13}}
\end{figure*}

\begin{figure*}
\center
\begin{verbatim}
static __uint128_t x, weyl, s; // s must be odd 

bool Collatz_Weyl(void){
    if(x % 2 == 1){
    x = (3*x+1)/2;}
    else{
    x = x/2;}
    x ^= (weyl += s);
    return x & 1;
}
\end{verbatim}
\caption{Implementation of the Collatz-Weyl Generator based on the original "3x+1" sequences in C++.\label{fig:14}}
\end{figure*}

Let us consider the simplest Collatz-Weyl Generator with $m=2$ that is based on the original Collatz sequences, as shown in Figure~\ref{fig:13} (or, equivalently, presented in Figure~\ref{fig:14}). This generator outputs only one bit per call, but it becomes very difficult for an attacker who does not know $s$, $weyl$, and other bits of $x$ to predict the next or past outputs of the generator after skipping the first few outputs. This is not due only to the non-invertibility of the mapping itself (to iterate back, we have to guess at least one bit at every step even if we know all the states of the generator). The least significant bits of such sequences generated by the proposed algorithm exhibit deterministic chaos, similar to that observed in Collatz sequences and their generalizations \cite{kontorovich2010stochastic,lagarias2023ultimate,
xu2019pseudo,givens2006conway,lagarias20103x+}, and thus passed all stringent tests of randomness. This raises the prospect of the possibility of constructing efficient cryptographic primitives based on such schemes---for example, an ultra-lightweight stream cipher. Note that if we define the states of the proposed generator on registers of unlimited size, it seems we obtain a generator with an infinite period. However, we set aside this issue as a hypothesis.

An even more complicated set of behaviors was exhibited by the variant presented in Figures~\ref{fig:15} and \ref{fig:16}. Multiplication may not be a suitable operation for cryptographic schemes as it might not be possible to execute it in constant time in some representations, and this renders the system vulnerable to timing attacks. Replacing it with multiplication in Galois fields appears to be a promising approach.

In future research, we plan to focus on the resistance of these generators to backtracking under two or more conditions\footnote{Variants with $m=1$ appear to be directly vulnerable to rotational cryptanalysis.} ($m \ge 2$) as well as their application to hashing functions,  message authentication codes, and ciphers.

\begin{figure*}
\center
\begin{verbatim}
static __uint128_t x, a, weyl, s; // s must be odd

__uint128_t CWG128_2(void){
      x = (-(x & 1) & (x * ((a += x) >> 1)) | ~- (x & 1) & ((x >> 1) * (a | 1))) ^ (weyl += s);
    return a >> 96 ^ x;
}
\end{verbatim}
\caption{Implementation of the CWG128-2 generator in C++.\label{fig:15}}
\end{figure*}

\begin{figure*}
\center
\begin{verbatim}
static inline uint64_t rotr(const uint64_t n, int k) {
    return (n >> k) | (n << (64 - k));
}

static uint64_t x, a, weyl, s; // s must be odd

uint64_t CWG64_rot_2(void){
    a += x;
    x = (-(x & 1) & (x * rotr(a >> 1,a >> 58)) | ~- (x & 1)
    & (rotr(x >> 1, x >> 58) * (a | 1))) ^ (weyl += s);
    return rotr(a, 48) ^ x;
}
\end{verbatim}
\caption{Implementation of the CWG64-rot-2 generator in C++.\label{fig:16}}
\end{figure*}

\section*{Acknowledgements}

I would like to thank professor Dan Tamir, with whom I established collaboration in 2021 and whose article \cite{xu2019pseudo} inspired me to develop my ideas. During the two years of our collaboration, the schemes I worked on evolved and were significantly modified by me under the pressure of Dan's critical comments. 

\newpage
\bibliographystyle{elsarticle-num}
\bibliography{References}

\begin{thebibliography}{10}
\expandafter\ifx\csname url\endcsname\relax
  \def\url#1{\texttt{#1}}\fi
\expandafter\ifx\csname urlprefix\endcsname\relax\def\urlprefix{URL }\fi
\expandafter\ifx\csname href\endcsname\relax
  \def\href#1#2{#2} \def\path#1{#1}\fi

\bibitem{von1951various}
J.~Von~Neumann, et~al., Various techniques used in connection with random
  digits, Applied Math Series 12~(36-38) (1951) 1.

\bibitem{fischer2012backtracking}
M.~J. Fischer, M.~Paterson, E.~Syta, On backtracking resistance in pseudorandom
  bit generation, Tech. rep., Technical Report TR-1466, Yale, October 2012.
  http://cs. yale. edu~… (2012).

\bibitem{thakkar2016additive}
A.~Thakkar, M.~Jagadale, Additive collatz trajectories, arXiv preprint
  arXiv:1611.03919 (2016).

\bibitem{conway1972unpredictable}
J.~Conway, Unpredictable iterations, The Ultimate Challenge: The 3x+ 1 Problem
  (1972) 49--52.

\bibitem{kurtz2007undecidability}
S.~A. Kurtz, J.~Simon, The undecidability of the generalized collatz problem,
  in: TAMC, Vol.~7, Springer, 2007, pp. 542--553.

\bibitem{marsaglia2003xorshift}
G.~Marsaglia, Xorshift rngs, Journal of Statistical software 8 (2003) 1--6.

\bibitem{widynski2017middle}
B.~Widynski, Middle-square weyl sequence rng, arXiv preprint arXiv:1704.00358
  (2017).

\bibitem{fog2015pseudo}
A.~Fog, Pseudo-random number generators for vector processors and multicore
  processors, Journal of modern applied statistical methods 14~(1) (2015) 23.

\bibitem{flajolet1989random}
P.~Flajolet, A.~M. Odlyzko, Random mapping statistics, in: Workshop on the
  Theory and Application of of Cryptographic Techniques, Springer, 1989, pp.
  329--354.

\bibitem{wiki:birthday}
{Birthday problem},
  \href{https://en.wikipedia.org/wiki/Birthday_problem}{Birthday problem ---
  {W}ikipedia{,} the free encyclopedia}, [Online; accessed 04-July-2023]
  (2023).
\newline\urlprefix\url{https://en.wikipedia.org/wiki/Birthday_problem}

\bibitem{caflisch1998monte}
R.~E. Caflisch, Monte carlo and quasi-monte carlo methods, Acta numerica 7
  (1998) 1--49.

\bibitem{larcher1998random}
P.~H.~G. Larcher, P.~Hickernell, P.~L'Ecuyer, H.~Niederreiter, S.~Tezuka,
  C.~Xing, Random and quasi-random point sets, Vol. 138, Springer Science \&
  Business Media, 1998.

\bibitem{gentle2003random}
J.~E. Gentle, Random number generation and Monte Carlo methods, Vol. 381,
  Springer, 2003.

\bibitem{o2015pcg}
M.~O’Neill, Pcg, a family of better random number generators, PCG is a simple
  and fast statistically good PRNG (2015).

\bibitem{steele2014fast}
G.~L. Steele~Jr, D.~Lea, C.~H. Flood, Fast splittable pseudorandom number
  generators, ACM SIGPLAN Notices 49~(10) (2014) 453--472.

\bibitem{matsumoto2007common}
M.~Matsumoto, I.~Wada, A.~Kuramoto, H.~Ashihara, Common defects in
  initialization of pseudorandom number generators, ACM Transactions on
  Modeling and Computer Simulation (TOMACS) 17~(4) (2007) 15--es.

\bibitem{blackman2021scrambled}
D.~Blackman, S.~Vigna, Scrambled linear pseudorandom number generators, ACM
  Transactions on Mathematical Software (TOMS) 47~(4) (2021) 1--32.

\bibitem{rukhin2001statistical}
A.~Rukhin, J.~Soto, J.~Nechvatal, M.~Smid, E.~Barker, A statistical test suite
  for random and pseudorandom number generators for cryptographic applications,
  Tech. rep., Booz-allen and hamilton inc mclean va (2001).

\bibitem{brown2018dieharder}
R.~G. Brown, D.~Eddelbuettel, D.~Bauer, Dieharder, Duke University Physics
  Department Durham, NC (2018) (2018) 27708--0305.

\bibitem{l2007testu01}
P.~L'ecuyer, R.~Simard, Testu01: Ac library for empirical testing of random
  number generators, ACM Transactions on Mathematical Software (TOMS) 33~(4)
  (2007) 1--40.

\bibitem{doty10practically}
C.~Doty-Humphrey, Practically random: C++ library of statistical tests for
  rngs. version 0.94. 2018, URl: http://pracrand. sourceforge. net (cit. on pp.
  10, 13).

\bibitem{too-big-to-fail}
M.~O'Neill, Too big to fail,
  \url{https://www.pcg-random.org/posts/too-big-to-fail.html}, accessed:
  2023-06-27 (2017).

\bibitem{pcg-passes-practrand}
M.~O'Neill, Pcg passes practrand,
  \url{https://www.pcg-random.org/posts/pcg-passes-practrand.html}, accessed:
  2023-06-27 (2017).

\bibitem{salmon2011parallel}
J.~K. Salmon, M.~A. Moraes, R.~O. Dror, D.~E. Shaw, Parallel random numbers: as
  easy as 1, 2, 3, in: Proceedings of 2011 international conference for high
  performance computing, networking, storage and analysis, 2011, pp. 1--12.

\bibitem{tan2021thundering}
H.~Tan, X.~Chen, Y.~Chen, B.~He, W.-F. Wong, Thundering: generating multiple
  independent random number sequences on fpgas, in: Proceedings of the ACM
  International Conference on Supercomputing, 2021, pp. 115--126.

\bibitem{li2013software}
Y.~Li, P.~Chow, J.~Jiang, M.~Zhang, S.~Wei, Software/hardware parallel
  long-period random number generation framework based on the well method, IEEE
  Transactions on Very Large Scale Integration (VLSI) Systems 22~(5) (2013)
  1054--1059.

\bibitem{birthdaytest}
M.~O'Neill, A birthday test: Quickly failing some popular prngs, \url{A
  Birthday Test: Quickly Failing Some Popular PRNGs}, accessed: 2023-07-05
  (2017).

\bibitem{ROOT}
{ROOT} data analysis framework, \url{https://root.cern}, accessed: 2023-06-27
  (2023).

\bibitem{kontorovich2010stochastic}
A.~V. Kontorovich, J.~C. Lagarias, Stochastic models for the 3x+ 1 and 5x+ 1
  problems and related problems, The Ultimate Challenge: The 3x+ 1 Problem
  (2010) 131--188.

\bibitem{lagarias2023ultimate}
J.~C. Lagarias, The Ultimate Challenge: The $3 x+ 1$ Problem, American
  Mathematical Society, 2023.

\bibitem{xu2019pseudo}
D.~Xu, D.~E. Tamir, Pseudo-random number generators based on the collatz
  conjecture, International Journal of Information Technology 11~(3) (2019)
  453--459.

\bibitem{givens2006conway}
R.~M. Givens, On conway's generalization of the 3x+ 1 problem (2006).

\bibitem{lagarias20103x+}
J.~C. Lagarias, The 3x+ 1 problem: An overview, American Mathematical Society
  (2010).

\end{thebibliography}

\end{document}